**Room temperature magnetic entropy change and magnetoresistance in $La_{0.70}(Ca_{0.30-x}Sr_x)MnO_3$:Ag 10% ($x$ = 0.0-0.10)**

R. Jha, Shiva Kumar Singh,[*] Anuj Kumar and V.P.S Awana[†]

Quantum Phenomena and Applications, National Physical Laboratory (CSIR), New Delhi-110012, India

Abstract

The magnetic and magnetocaloric properties of polycrystalline $La_{0.70}(Ca_{0.30-x}Sr_x)MnO_3$:Ag 10% manganite have been investigated. All the compositions are crystallized in single phase orthorhombic *Pbnm* space group. Both, the Insulator-Metal transition temperature ($T^{IM}$) and Curie temperature ($T_c$) are observed at 298 K for $x$ = 0.10 composition. Though both $T^{IM}$ and $T_c$ are nearly unchanged with Ag addition, the *MR* is slightly improved. The *MR* at 300 K is found to be as large as 31% with magnetic field change of 1Tesla, whereas it reaches up to 49% at magnetic field of 3Tesla for $La_{0.70}Ca_{0.20}Sr_{0.10}MnO_3$:$Ag_{0.10}$ sample. The maximum entropy change ($\Delta S_{Mmax}$) is 7.6 J.Kg$^{-1}$.K$^{-1}$ upon the magnetic field change of 5Tesla, near its $T_c$ (300.5 K). The $La_{0.70}Ca_{0.20}Sr_{0.10}MnO_3$:$Ag_{0.10}$ sample having good MR (31%$^{1Tesla}$, 49%$^{3Tesla}$) and reasonable change in magnetic entropy (7.6 J.Kg$^{-1}$.K$^{-1}$, 5 Tesla) at 300 K can be a potential magnetic refrigerant material at ambient temperatures.

**Introduction**

Manganites are considered to be promising candidate for the technological applications such as bolometer and magnetic refrigeration [1-7]. Along with the all other fascinating properties, the presence of magnetocaloric nature makes them more outstanding material. Practically one desires to have higher Temperature coefficient of resistance (*TCR*), *MR* and magnetic entropy change near room temperature i.e. at around 300 K. It is seen that the maximum *MR* as well as *TCR* in hole-doped manganites occur near the insulator-metal (*IM*) transition $T^{IM}$ being accompanied with ferromagnetic (*FM*)-paramagnetic (*PM*) transition (Curie Temperature $T_c$). The steep transition about *IM* crossover determines the sensitivity as well as active zone for these sensors. Since magnetic refrigeration has a lot of advantages over gas refrigeration, manganites have been getting attention [4, 6-7]. Also, since the magnetic properties of perovskite manganites, Curie temperature and saturation magnetization, are strongly doping-dependent, these typical materials are believed to be good candidates for magnetic refrigeration at various temperatures [5-11].

The magnetocaloric effect (*MCE*) is an isothermal magnetic entropy change or an adiabatic temperature change ($\Delta T_{ad}$) of a magnetic material caused by an applied magnetic field. The adiabatic temperature change $\Delta T_{ad}$ is mainly affected by the magnetic entropy change $|\Delta S_M|$ [12]. The magnetic entropy change $|\Delta S_M|$ induced by a magnetic field change is another important parameter to describe the magnetocaloric effect. A constant magnetic entropy change over the working temperature range is required in an ideal Ericsson refrigeration cycle [13]. It has been observed that heavy rare-earth and their compounds are good candidates for finding a large *MCE*, due to their large magnetic moments [14, 15]. The highest *MCE* involving a second-order transition is found in gadolinium, which can be used to achieve cooling between 270 and 310 K [14]. However, the cost of a magnetic refrigerant using gadolinium is quite expensive, which limits the usage of it as an active magnetic refrigerant (*AMR*) in magnetic refrigerators. Further efforts to investigate new materials exhibiting large *MCE* in relatively low applied field are of significant importance [11, 15-16]. An *AMR* material should have large magnetic entropy change induced by low magnetic field change. However higher resistivity of manganites is favorable for reducing eddy current heating though their maximum entropy change is smaller than rare earth compounds.

Various attempt has been made to increase *TCR* and *MR* in $La_{2/3}Ca_{1/3}MnO_3:Ag_y$ composites [17-18]. Higher *TCR* with optimized reasonable *MR* at low fields is seen below room temperature (< 265 K). Some of us found 30% *MR* at 1 Tesla and *TCR* as high as 9 %/K above 300K in $La_{0.70}Ca_{0.20}Sr_{0.10}MnO_3:Ag_{0.2}$ [5]. These values are quite reasonable in bulk polycrystalline samples and can be used in the bolometric and infrared detectors. However there are scant reports on LCMO:Ag composites study for magnetic refrigeration or for magneto-caloric applications. In this report we have studied $La_{0.70}Ca_{0.30-x}Sr_xMnO_3:Ag_{0.10}$ composites for *MCE* and applications. Our results show that Ag addition improves *MR* and magneto-entropy change. The occurrence of maximum entropy change ($\Delta S_{Mmax}$) near its $T_c$ (300.5 K) makes it much of practical importance and it can be used for magnetic refrigeration.

**Experimental**

The samples are synthesized in air by solid-state reaction route. The stoichiometric mixture of $La_2O_3$, $SrCO_3$, CaO and $MnO_2$ are ground thoroughly, calcined at 1000ºC for 12h and then pre-sintered at 1100 ºC, 1200 ºC, 1300 ºC and 1400 ºC for 20h with intermediate grindings. Finally, the powders are palletized and sintered at 1420 ºC for 20h in air. Samples are cooled very slowly ($1^0$C /minute) from 1420ºC to room temperature. In Ag composite samples $Ag_2O$ is mixed by weight percentage before final sintering. The phase formation is checked for each sample with powder diffractometer, Rigaku (Cu-Kα radiation) at room temperature. The phase purity analysis and lattice parameter refining are performed by Rietveld refinement programme (Fullprof version). The resistivity and magnetization measurements of all samples are carried out applying a field magnitude up to 5 T using Physical Properties Measurement system- Quantum Designed PPMS-14 T.

1. **Results and discussion**

All the samples are crystallized in single phase [Fig. 1(a) and (b)]. This is confirmed from the Rietveld analysis of powder X-ray diffraction patterns. All the compositions $La_{0.70}Ca_{0.30-x}Sr_xMnO_3$ ($x$ = 0.0, 0.05 and 0.10) are fitted in orthorhombic *Pbnm* space group. Ionic radii of $Ca^{2+}$ with coordination number (CN) VI is 1.0 Å and the ionic radii of $Sr^{2+}$ with CN VI is 1.18 Å [19]. Increase in lattice parameters indicates that substitution by Sr at Ca site. Fitted parameters are shown in table 1. In Ag composite samples, it is possible for Ag to enter the manganite lattice, which would increase the $Mn^{4+}$ content [20]. Also, Ag is volatile above 1000 $^0$C and it may restrict presence of Ag less than the desired stoichiometry. To avoid these, $Ag_2O$ is mixed in final sintering. The Rietveld refinement of Ag added samples shows that presence of Ag peak in [see Fig. 1(b)]. This clearly indicates that most of Ag is present at the grain boundary.

Fig. 2 depicts *RTH* plot of $La_{0.70}Ca_{0.30}MnO_3:Ag_{0.10}$ composition up to applied field of 7 T, which shows transition near 270 K in zero field. With increasing field transition shifts towards higher temperature, which is attributed to enhancement in the ferromagnetic interactions with higher applied fields. Inset of Fig. 2 shows *MR* of the same. A maximum 58 % change of *MR* can be seen with the change in applied field of 3 T at 270 K. Although it has better *MR* but the working temperature is far below room temperature. Earlier it is reported [5, 7, 11] with doping of Sr at Ca site the transition temperature increases and it is found around 308 K for Sr = 0.10. Although the $T_c$ increases with Sr doping but it is at the cost of sharpness of transition. However, synthesis condition and oxygen content, largely determine the $T_c$ and sharpness of transition [21-22]. In other reports [5, 11], $T_c$ was found around 306-308 K with Sr = 0.10. In the studied samples the $T_c$ is found to be 298 K with Sr content of 0.10. This is determined through derivative of the magnetization data (*M-T*) in an applied field of 0.1 T [Fig. 3]. The observed difference in $T_c$ may be due to deficiency of oxygen content. In earlier report [5] samples are oxygen annealed at 1200 C after being synthesized at 1400 C, whereas the studied samples are slowly cooled in air from 1420 C to room temperature. In any case working temperature of 298K is good enough for room temperature applications.

The large magnetic entropy change in manganites mainly originates from the considerable variation of magnetization near $T_c$. Also, the spin lattice coupling in the magnetic ordering process plays an important role [8-11]. It is reported that with Ag addition in manganites, though sharpness of transition improves, the $T^{IM}$ remains nearly invariant [7, 23-24]. Fig. 4a and 4b show the *MR* of $La_{0.70}Ca_{0.20}Sr_{0.10}MnO_3$ and $La_{0.70}Ca_{0.20}Sr_{0.10}MnO_3:Ag_{0.10}$ composition respectively. It can be seen that maximum *MR* increases with Ag addition. The MR at 3 Tesla is 43 % and 49 % in Ag free and Ag added samples respectively. Contrasting interpretations are argued about the mechanism of *MR* increase with Ag addition. Low melting point silver segregates at the grain boundaries and lead to reduction in intergrain tunnel resistance [5, 7]. Intergrain region offers more resistance in conduction as it behaves as non metallic and nonmagnetic region. Ag addition provides a conducting channel between the grains which leads to sharper transition. It is reported that Ag addition improves FM in nonmagnetic intergrain region [23]. A remarkable improvement in the magnetic homogeneity is indicated by narrower ferromagnetic resonance (*FMR*) line widths in thin films [23]. Thereby, a more abrupt reduction of magnetization occurs, which results in a significant magnetic-entropy change near $T_c$ and thus a better *MCE* can be obtained in manganites. In addition to this it is argued that Mn spin disorders occur at the phase interfaces due to which Mn-Mn magnetic exchange is interrupted as Ag segregates at the grain boundaries [7]. This magnetic inhomogeneity leads to the increase in resistivity [25-27]. With an external field applied, spin scattering is suppressed and thus enhanced *MR* is obtained. Considering the points given above, it is thus assumed that increase in MR and sharpness of transition at $T_c/T_{MI}$ is obvious for Ag added sample.

Considering the fact that better *MR* is observed in Ag added sample and we are interested in near room temperature, the isothermal magnetization is done for $La_{0.70}Ca_{0.20}Sr_{0.10}MnO_3:Ag_{0.10}$ sample. The magnetocaloric effect can be measured either by the adiabatic change of temperature under the application of a magnetic field or through the measurement of isothermal magnetization versus field at different temperatures [28]. We used the second one to avoid complications related with adiabatic measurement. Fig. 5 shows the representative isothermal curves of magnetization with the applied field up to 5 T for $La_{0.70}Ca_{0.20}Sr_{0.10}MnO_3:Ag_{0.10}$ sample. Isothermal curves are obtained around $T_c$ from 290 to 310 K at an interval of 1 and 2 K. The change in magneto-entropy is related to the magnetization by the Maxwell relation [28].

This equation can be rewritten as

If the magnetization measurement is done at small temperature interval and discrete fields, this equation can be approximated as

Here, $M_i$ and $M_{i+1}$ are the magnetization values measured at temperatures $T_i$ and $T_{i+1}$ in an applied field $H$, respectively. Thus the magneto-entropy change can be calculated through isothermal magnetization curves.

Fig. 6 shows the magnetic entropy change $|\Delta S_M|$ as a function of temperature for the $La_{0.70}Ca_{0.20}Sr_{0.10}MnO_3$:$Ag_{0.10}$ sample with different magnetic field changes $\Delta H$. In the same magnetic field change, $|\Delta S_M|$ variation with temperature shows a peak near $T_c$ (300.5 K). Upon the change of 5 Tesla applied field, the highest value of $|\Delta S_M|$ is 7.6 J.Kg$^{-1}$.K$^{-1}$. With the same field change (5 Tesla) the value of $|\Delta S_M|$ is 7.45 J.Kg$^{-1}$.K$^{-1}$ was found in single crystal of $La_{0.70}Ca_{0.20}Sr_{0.10}MnO_3$ [7]. It means Ag addition provides similar results to that of single crystals of non Ag added samples. At commercial level it is easy to synthesize bulk material than single crystals. Thus Ag composites of bulk manganites could be the potential candidate for magnetic refrigeration.

**Conclusion**

Improvement in magnetic and magnetocaloric properties has been observed with Ag addition in polycrystalline manganites. Both, the Insulator-Metal transition and Curie temperatures areo observed at 298 K for $x = 0.10$ composition. While $T^{IM}$ remains constant with Ag addition, ~ 6 % increase of *MR* is observed with same. The *MR* at 300 K is found to be as large as 31% with magnetic field change of 1T in Ag added $x = 0.10$ composition. The maximum entropy change ($\Delta S_{Mmax}$) is found to be 7.6 J.Kg$^{-1}$.K$^{-1}$ upon the magnetic field change of 5T, near its $T_c$. It shows that Ag addition is competent to that of single crystals of non Ag added samples. Thus Ag composites of bulk polycrystalline manganites could be the potential candidate for magnetic refrigeration.

**Acknowledgements**

The authors would like to thank DNPL Prof. R. C. Budhani for his constant support and encouragement. Authors S. K. Singh and A. Kumar would like to acknowledge *CSIR*, India, for providing fellowships.

**References**

[1] R. von Helmolt, J. Wocker, B. Holzapfel, M. Schultz, and K. Samver Phys. Rev. Lett. 71 (1993) 2331
[2] S. Jin, T. H. Tiefel, M. McCormack, R. A. Fastnacht, R. Ramesh, and L. H. Chen, Science 264 (1994) 413
[3] A.-M. Haghiri-Gosnet and J.-P. Renard, J. Phys. D 36 (2003) R127
[4] A. F. Lacaze, R. Beranger, G. Bon Mardion, G. Claudet, A. A. Lacaze Cryogenics 23 (1983) 427
[5] V. P. S. Awana, Rahul Tripathi, Neeraj Kumar, H. Kishan, G. L. Bhalla, R. Zeng, L. S. Sharth Chandra, V. Ganesan, and H. U. Habermeier J. Appl. Phys 107 (2010) 09D723
[6] G. C. Lin, X. L. Yu, Q. Wei, J. X. Zhang Materials Letters 59 (2005) 2129

[7] Y-H Huang, C-H Yan, F. Luo, W. Song, Z-M. Wang, and C-S Liao Appl. Phys. Lett. 81 (2002) 76

[8] Z. B. Guo, Y. W. Du, J. S. Zhu, H. Huang, W. P. Ding, and D. Feng, Appl. Phys. Lett. 78 (1997) 1142

[9] X. Bohigas, J. Tejada, E. del Barco, X. X. Zhang, and M. Sales, Appl. Phys. Lett. 73 (1998) 390

[10] A. Szewczyk, H. Szymczak, A. Wisniewski, K. Piotrowski, R. Kartaszynski, B. Dabrowski, S. Kolesnik, and Z. Bukowski, Appl. Phys. Lett. 77 (2000) 1026

[11] M. H. Phan, S. C. Yu, and N. H. Hur, Appl. Phys. Lett. 86 (2005) 72

[12] X. Bohigas, J. Tejada, M. L. Marı́nez-Sarrió́n, S. Tripp, R. Black, J. Magn. Magn. Mater. 208 (2001) 85

[13] T. Hashimoto, T. Kuzuhara, M. Sahashi, K. Inomata, A. Tomokiyo, H.Yayama, J. Appl. Phys. 62 (1987) 3873

[14] R. D. McMichael, J.J. Ritter, R.D. Shull, J. Appl. Phys. 73 (1993) 6946

[15] S. Yu. Dan'kov, et al., Phys. Rev. B 57 (1998) 3478

[16] V. K. Pecharsky, K. A. Gschneidner Jr., Phys. Rev. Lett. 78 (1997) 4494.

[17] H. Wada and Y. Tanabe, Appl. Phys. Lett. 79 (2001) 3302

[18] Tegus, E. Bruck, K. H. J. Buschow, and F. R. de Boer, Nature (London) 415 (2002) 150

[19] R. D. Shannon, Acta Crystallogr. Sect. A: Cryst. Phys. Diffr. Theor. Gen. Crystallogr. A32, 751 1976

[20] T. Tang, Q. Q. Cao, K. M. Gu, H. Y. Xu, S. Y. Zhang, and Y. W. Du, Appl. Phys. Lett. 77 (2000) 723

[21] B C Nam, W S Kim, H S Choi, J C Kim, N H Hur, I S Kim and Y K Park J. Phys. D Appl. Phys. 34 (2001) 54–59

[22] R. Tripathi, V. P. S. Awana, N. Panwar, G. L. Bhalla, H. U. Habermier, S. K. Agarwal, and H. Kishan, J. Phys. D 42 (2009) 175002

[23] R. Shreekala, M. Rajeshwari, S. P. Pai, S. E. Lofland, V. Smolyaninova, K. Ghosh, S. B. Ogale, S. M. Bhagat, M. J. Downes, R. L. Greene, R. Ramesh, and T. Venkatesan, Appl. Phys. Lett. 74 (1999) 2857

[24] V. P. S. Awana, R. Tripathi, S. Balamurugan, H. Kishan, and E. Takayama-Muromachi, Solid State Commun. 140 (2006) 410

[25] L. Balcells, A. E. Carrillo, B. Martinez, and J. Fontcuberta, Appl. Phys. Lett. 74 (1999) 4014

[26] D. K. Petrov, L. Krusin-Elbaum, J. Z. Sun, C. Field, and P. R. Duncombe, Appl. Phys. Lett. 75 (1999) 995

[27] S. Gupta, R. Ranjit, C. Mitra, P. Raychaudhuri, and R. Pinto, Appl. Phys. Lett. 78 (2001) 362

[28] A. H. Morrish, *The Physical Principles of Magnetism* IEEE, New York, (2001)

**Figure Caption**

Fig. 1: Rietveld fitted *XRD* pattern of (a) $La_{0.70}(Ca_{0.30-x}Sr_x)MnO_3$ ($x = 0.00$ and $0.05$); (b) $La_{0.70}(Ca_{0.30-x}Sr_x)MnO_3$ ($x = 0.10$) and $La_{0.70}(Ca_{0.30-x}Sr_x)MnO_3$:Ag ($x = 0.10$) with space group *Pbnm*. In Fig 1(b) * represents peaks associated with Ag.

Fig. 2: *RTH* plot of $La_{0.70}Ca_{0.30}MnO_3$:$Ag_{0.10}$ composition up to applied field of 7 T. Inset of Fig. 2 shows *MR* of the same at various temperature.

Fig. 3: Magnetization (*M-T*) and its derivative in an applied field of 0.1 T of the $La_{0.70}(Ca_{0.30-x}Sr_x)MnO_3$:Ag ($x = 0.10$) sample.

Fig. 4: *MR* of (a) $La_{0.70}Ca_{0.20}Sr_{0.10}MnO_3$ and (b) $La_{0.70}Ca_{0.20}Sr_{0.10}MnO_3$:$Ag_{0.10}$ composition with field at various temperatures.

Fig. 5: Representative isothermal curves of magnetization with the applied field up to 5 T for $La_{0.70}Ca_{0.20}Sr_{0.10}MnO_3$:$Ag_{0.10}$ sample. Isothermal curves are obtained around $T_c$ from 290 to 310 K at an interval of 1 and 2 K.

Fig. 6: The magnetic entropy change $|\Delta S_M|$ as a function of temperature for the $La_{0.70}Ca_{0.20}Sr_{0.10}MnO_3$:$Ag_{0.10}$ sample with different magnetic field changes $\Delta H$.

Table 1: Rietveld Refined lattice parameters and unit cell volume of $La_{0.70}(Ca_{0.30-x}Sr_x)MnO_3$ ($x = 0.00, 0.05$ and $0.10$).

| $La_{0.70}(Ca_{0.30-x}Sr_x)MnO_3$ | $R_p$ | $R_{wp}$ | Chi$^2$ | a (Å) | b (Å) | c (Å) | Vol. (Å$^3$) |
|---|---|---|---|---|---|---|---|
| $x = 0.00$ | 4.73 | 6.03 | 2.34 | 5.45 (4) | 5.47 (1) | 7.70 (9) | 230.06 (6) |
| $x = 0.05$ | 4.87 | 6.20 | 2.08 | 5.45 (8) | 5.48 (2) | 7.71 (1) | 230.79 (6) |
| $x = 0.10$ | 4.92 | 6.24 | 2.61 | 5.46 (4) | 5.49 (5) | 7.72 (2) | 231.92 (1) |

**Fig. 1(a)**

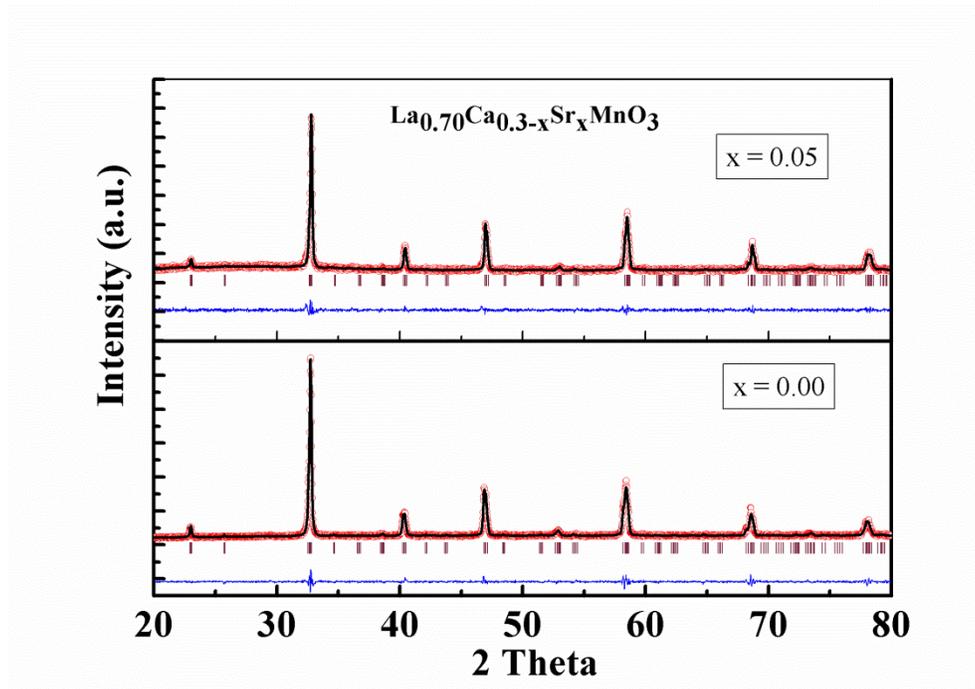

**Fig. 1(b)**

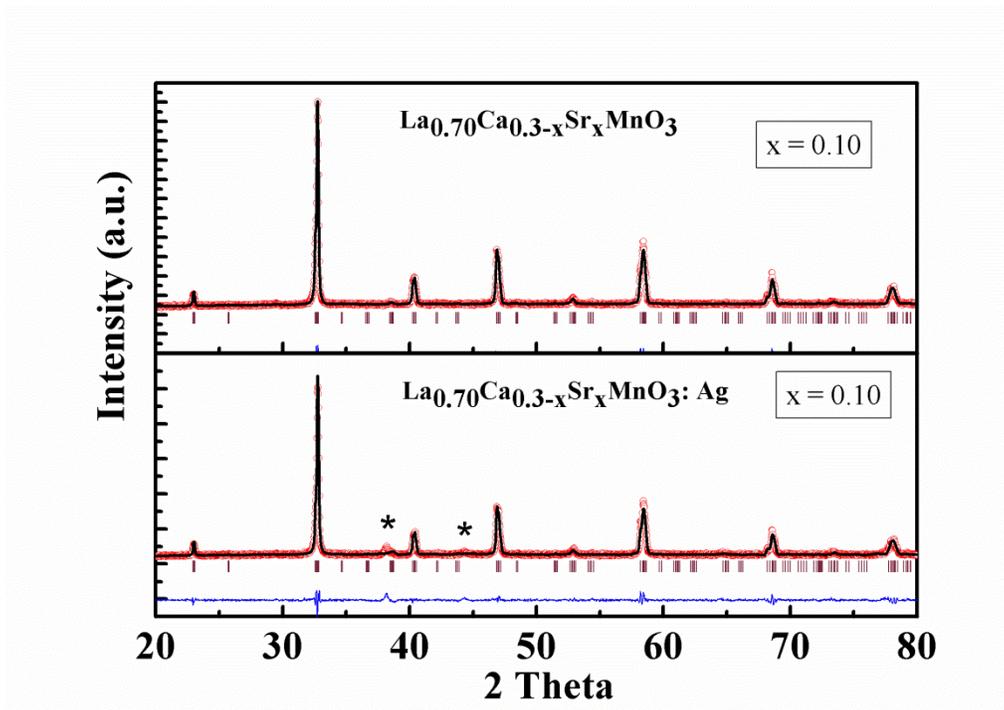

**Fig. 2**

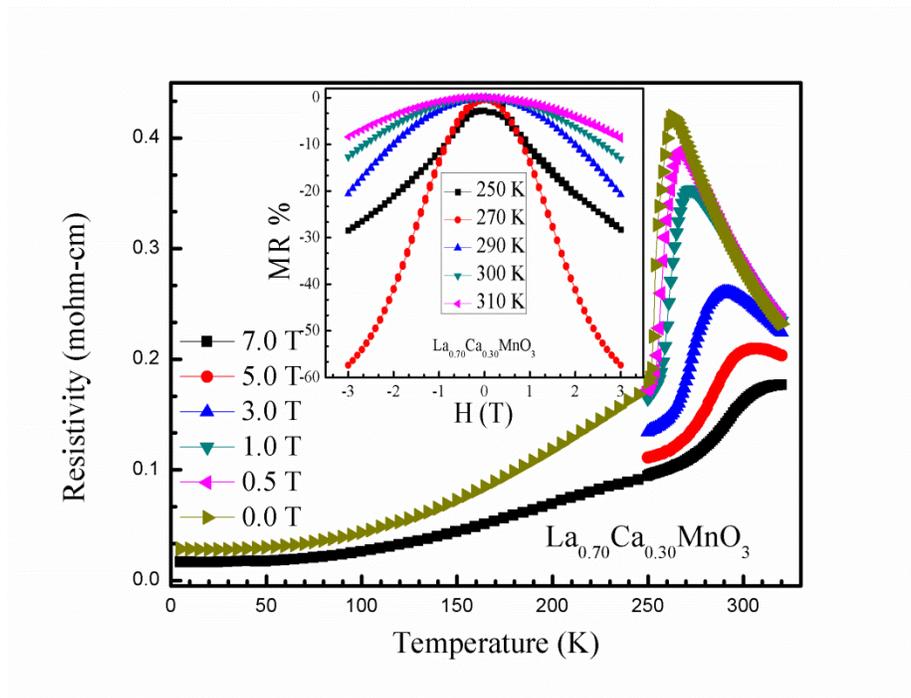

**Fig.3**

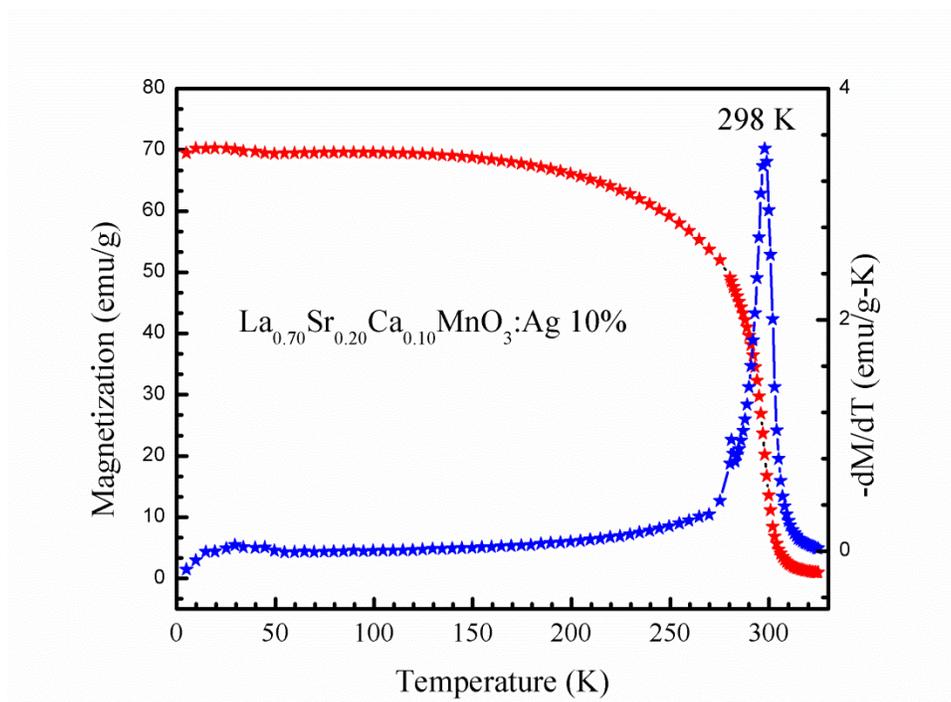

**Fig. 4(a)**

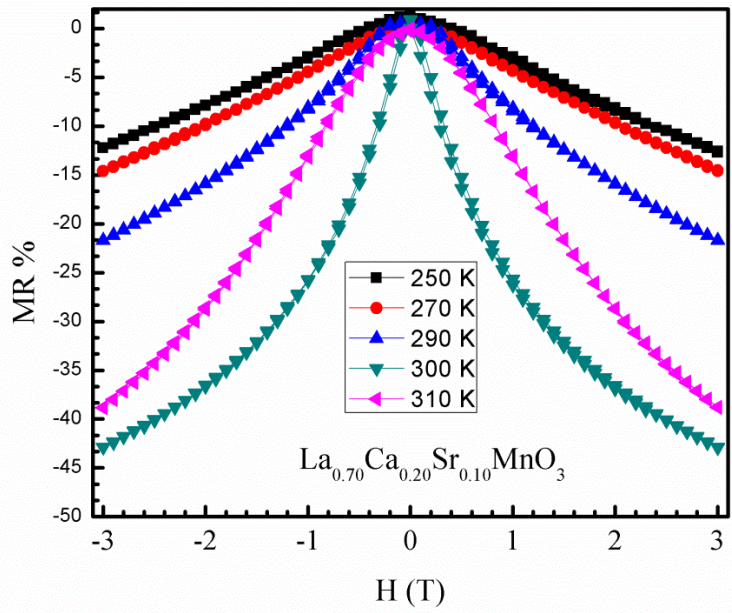

**Fig. 4(b)**

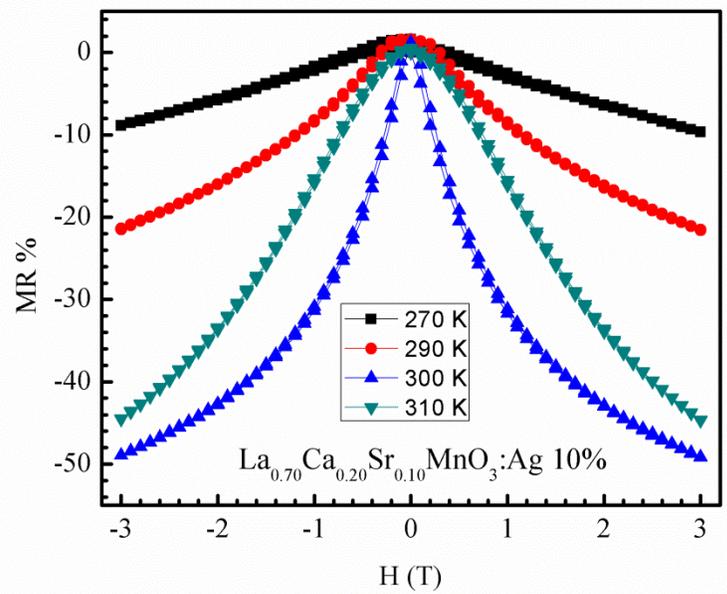

**Fig. 5**

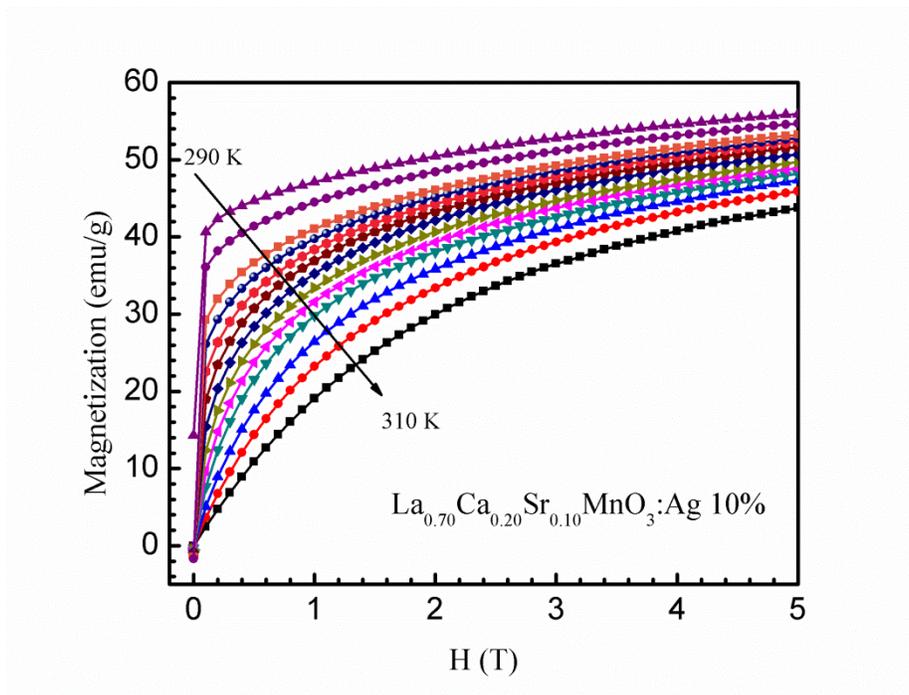

**Fig.6**

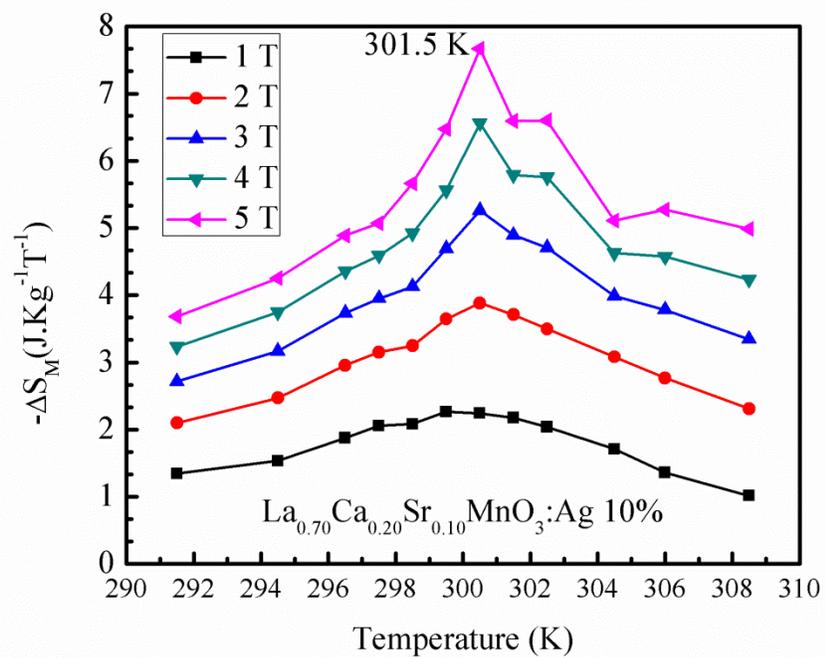